\documentclass[fleqn,usenatbib]{mnras}

\usepackage{newtxtext,newtxmath}
\usepackage[T1]{fontenc}

\usepackage[normalem]{ulem} 

\DeclareRobustCommand{\VAN}[3]{#2}
\let\VANthebibliography\thebibliography
\def\thebibliography{\DeclareRobustCommand{\VAN}[3]{##3}\VANthebibliography}

\newcommand{\teff}{T_{\rm eff}}
\newcommand{\logg}{\log g}
\newcommand{\vmic}{\xi_{\rm t}}
\newcommand{\vmac}{V_{\rm mac}}

\usepackage{graphicx}	% Including figure files
\usepackage{amsmath}	% Advanced maths commands
\title[Diffusion in AI Phe]{Evidence for elemental diffusion in the eclipsing binary star AI Phoenicis\thanks{Based on observations made with ESO Telescopes at the La Silla Paranal Observatory under programme ID 0106.D-0165(A), and data obtained from the ESO Science Archive Facility with DOIs: https://doi.org/10.18727/archive/33. }
}
\author[P. F. L. Maxted et al.]{
Pierre F. L. Maxted,$^{1\,\href{https://orcid.org/0000-0003-3794-1317}{\includegraphics[scale=0.5]{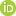}}}$ 
Nicholas Storm,$^{2\,\href{https://orcid.org/0000-0002-5259-3974}{\includegraphics[scale=0.5]{figs/orcid.jpg}}}$
Andreas J. Korn,$^{3\,\href{https://orcid.org/0000-0002-3881-6756}{\includegraphics[scale=0.5]{figs/orcid.jpg}}}$
Mar\'ilia Carlos,$^{4\href{https://orcid.org/0000-0003-1757-6666}{\includegraphics[scale=0.5]{figs/orcid.jpg}}}$
Matthew R. Gent,$^{5,2\,\href{https://orcid.org/0000-0002-5725-7160}{\includegraphics[scale=0.5]{figs/orcid.jpg}}}$
\newauthor
Sviatoslav B. Borisov,$^{6\,\href{https://orcid.org/0000-0002-2516-9000}{\includegraphics[scale=0.5]{figs/orcid.jpg}}}$
Paula Jofré,$^{7\,\href{https://orcid.org/}{\includegraphics[scale=0.5]{figs/orcid.jpg}}}$
Maria Bergemann,$^{2\,\href{https://orcid.org/0000-0002-9908-5571}{\includegraphics[scale=0.5]{figs/orcid.jpg}}}$
Hans-Günter Ludwig,$^{8\,\href{https://orcid.org/0000-0001-9333-4410}{\includegraphics[scale=0.5]{figs/orcid.jpg}}}$ and
\newauthor
Nicola J. Miller$^{3\,\href{https://orcid.org/0000-0001-9550-1198}{\includegraphics[scale=0.5]{figs/orcid.jpg}}}$\\
$^{1}$Astrophysics group, Keele University, Staffordshire, ST5 5BG, UK.\\
$^{2}$Max-Planck-Institut f\"{u}r Astronomie, K\"{o}nigstuhl 17, D-69117 Heidelberg, Germany.\\
$^{3}$Department of Physics and Astronomy, Uppsala University, Box 516, SE-75120 Uppsala, Sweden.\\
$^4$Observatório Nacional/MCTIC, R. Gen. José Cristino, 77, 20921-400, Rio de Janeiro, Brazil.\\
$^{5}$IRAP, Université de Toulouse, CNRS, CNES, 14, avenue Édouard Belin, F-31400, Toulouse, France.\\
$^{6}$Department of Astronomy, University of Geneva, Chemin Pegasi 51, 1290 Versoix, Switzerland.\\
$^{7}$Núcleo Milenio ERIS \& Instituto de Estudios Astrofísicos, Facultad de Ingeniería y Ciencias, Universidad Diego Portales,  Santiago, Chile.\\
$^{8}$Landessternwarte -- Zentrum für Astronomie, Universität Heidelberg, Königstuhl 12, 69117 Heidelberg, Germany.
}

\date{Accepted XXX. Received YYY; in original form ZZZ}

\pubyear{2023}

\begin{document}
\label{firstpage}
\pagerange{\pageref{firstpage}--\pageref{lastpage}}
\maketitle

% Abstract of the paper
\begin{abstract}
AI~Phe is an eclipsing binary star with an orbital period of 24.6\,days for which the surface gravity and effective temperature are known from direct measurements to very high precision and accuracy. 
We have obtained high-quality spectroscopy of the K0\,IV star during the total eclipse of the F7\,V companion, and also obtained spectra with a very high signal-to-noise ratio for this star and its F7\,V companion using the spectral disentangling technique. 
We have used these spectra to measure the abundances of iron and magnesium for both stars. 
We compare the values of [Fe/H] and [Mg/H] for the  F7\,V star and the K0\,IV star to stars in M67, an open cluster of similar age and metallicity to AI~Phe. 
We find that our [Fe/H] and [Mg/H] measurements clearly show the signature of elemental diffusion in the F7\,V star.
This suggests that AI~Phe can be used to test models of single stars that include diffusion and mixing of elements.
\end{abstract}

% Select between one and six entries from the list of approved keywords.
% Don't make up new ones.

%%%%%%%%%%%%%%%%%%%%%%%%%%%%%%%%%%%%%%%%%%%%%%%%%%

%%%%%%%%%%%%%%%%% BODY OF PAPER %%%%%%%%%%%%%%%%%%

\section{Introduction}
\label{sec:intro}
Precise and accurate stellar parameters allow us to put limits on poorly constrained aspects of stellar evolution. As solar-type stars evolve, their surface and core composition may undergo changes due to gravitational settling and other mixing process which are generally referred to as atomic diffusion, elemental diffusion or, simply, diffusion. This is expected both from very large diffusion signatures seen in certain classes of A-type stars in which surface convection is weak or absent \citep[][ and references therein]{Richer2000} as well as from stellar-structure calculations which include all effects of elemental diffusion \citep{Richard2002}.  

Diffusion studies among F-/G-/K-type stars were first performed on metal-poor globular clusters \citep{King1998, Gratton2001, Korn2007}. They ultimately revealed small but systematic abundance differences between stars located at the turn-off point (where convection is shallow and diffusion can run its course) and on the red-giant branch (in which the original composition of heavy elements is restored by deep convection). Such abundance differences are best described by models that self-consistently treat all the effects of elemental diffusion (including radiative levitation) moderated by an empirically determined amount of mixing at the bottom of the convective envelope \citep[][and references therein]{Richard2005} . 

For solar-metallicity F-/G-/K-stars, elemental diffusion has been investigated for about a decade \citep{Onehag2014}. All major spectroscopic surveys have studied the solar-age solar-metallicity open cluster M67 where effects are found to be of order 0.15\,dex in logarithmic abundance difference between turn-off point and the red-giant branch \citep{BertelliMota2018, Gao2018, Souto2019}. Not all abundance trends are fully understood \citep[see e.g. Mg in ][]{Souto2019} and the measurements may suffer from biases stemming from simplified modelling assumptions like one-dimensional atmospheres or local thermodynamic equilibrium (LTE) in line-formation calculations.
Another open cluster in which atomic diffusion trends have been traced in several elements is NGC~2420 with an age of around 2.5 Gyr \citep{2020A&A...643A.164S}.
Effectively removing stellar-parameter uncertainties, as in the case of AI Phe presented in this paper, can sharpen our view of the effects at play at the interface between stellar interiors and surface layers. 
 
AI~Phe is a bright eclipsing binary star ($V$=8.7) that has frequently been used as a benchmark for stellar evolution models since the mass and radius were measured with a precision of about 1\% by \citet{1988A&A...196..128A}.
\citet{2020MNRAS.498..332M} used the light curve of AI~Phe observed by the TESS mission \citep{2015JATIS...1a4003R} combined with the spectroscopic orbits of the two stars to obtain mass and radius measurements with a precision of about 0.1\% for both stars.
The accuracy of these estimates was ensured by using three independent sources for the spectroscopic orbits, and by comparing the results of several independent analyses of the light curve.
\citet{2020MNRAS.497.2899M} used these radius measurements combined with the parallax of the system and photometry from ultraviolet to infrared wavelengths to directly measure the effective temperature ($T_{\rm eff}$) of the two stars from their bolometric fluxes and angular diameters.

The very high quality of the mass, radius and effective temperature measurements for AI~Phe motivated us to obtain a high-quality spectrum for the K0\,IV subgiant star.
This is possible because its F7\,V companion it totally eclipsed for about 30 minutes during the primary eclipse. We have also applied our spectral disentangling technique to archival spectra of AI~Phe in order to obtain a high-quality spectrum of the F7V star.  
We have analysed these spectra to obtain precise estimates of [Fe/H] and [Mg/H] for both stars and find that these measurements provide evidence for depletion of heavy elements in the photosphere of the hotter component of this binary system.

\section{Observations}
\label{sec:obs}
\subsection{Spectroscopy}
We obtained four spectra of AI~Phe during the primary eclipse on the night 2020-10-27 using the UVES spectrograph on the European Southern Observatory (ESO) 8.2-m UT2 telescope (``Kueyen''). The exposure time per spectrum was 530\,s. Observations were obtained using both arms of the spectrograph with slit widths of 0.3\,arcsec on the red arm (resolving power $R\approx 100,000$) and 0.4\,arcsec on the blue arm ($R\approx 80,000$). 

We processed the raw science images and calibration data to produce 1-dimensional spectra using the version 6.1.8 of the UVES pipeline software\footnote{\url{https://www.eso.org/sci/software/pipelines/uves/uves-pipe-recipes.html}}  provided by ESO \citep{2000Msngr.101...31B} running within the EsoReflex environment \citep{2013A&A...559A..96F}.\footnote{\url{https://www.eso.org/sci/software/esoreflex/}} 
With the default settings used for the extraction there is a problem with the profile fitting step that results in poor quality data in orders 98\,--\,101 (602.4 \,--\,627.3\,nm). 
For these orders, we used the ``linear'' (non-optimal) extraction option to avoid this problem. 
This appears to be due to a spurious absorption feature at 616.0\,--\,616.5\,nm.
This option does not provide automatic cosmic-ray rejection but this is not a problem for our analysis because we have multiple exposures so these outliers are easy to identify and remove. 
The signal to noise ratio in the continuum near 550\,nm is approximately 150 per pixel per spectrum at a dispersion of 0.0014\,nm/pixel. 
The blue arm spectra cover the wavelength range 329\,--\,456\,nm and the red arm spectra cover the wavelength range 459\,--\,668\,nm. 
The data from each order were merged into a 1-dimensional spectrum using the algorithm described in  \citet{2023ApJS..266...11B}. 
This algorithm significantly reduces ``ripples'' in the spectra caused by inaccurate merging of the spectra from different echelle orders compared to the order-merging algorithm used in the UVES pipeline.

The ESO Science archive facility\footnote{\url{http://archive.eso.org}} contains 36 spectra of AI~Phe observed with the HARPS spectrograph on the ESO 3.6-m telescope \citep{2003Msngr.114...20M}. 
HARPS is a fibre-fed, cross-dispersed échelle spectrograph that delivers spectra covering the wavelength range 380\,--\,690\,nm at resolving power $R\approx 115,000$. 
These spectra were obtained between 2011-06-09 and 2017-12-11 using exposure times of 220\,--\,1200\,s. We used the reduced 1-dimensional spectra obtained by standard ESO pipeline processing for our analysis.

\subsection{Orbital ephemeris and flux contamination}
Two additional sectors of TESS photometry observed at a cadence of 120\,s have become available since the study by \cite{2020MNRAS.498..332M}. 
These observations span the date of our spectroscopic observations with UVES and so are useful to compute precisely the orbital phase at which these spectra were taken. 
We used {\sc lightkurve}\footnote{\url{https://docs.lightkurve.org/}}  \citep{2018ascl.soft12013L} to download the light curves of AI~Phe produced by the Science Processing Operations Centre (SPOC) from the Mikulski Archive for Space Telescopes (MAST).\footnote{\url{https://archive.stsci.edu/}} 
We used the SAP\_FLUX measurements in the files for our analysis. Inspection of the data for sector 29 clearly show that the background flux has been underestimated in the processing of these data. 
We subtracted a scaled version of the background flux estimate SAP\_BKG provided in the same file from the SAP\_FLUX values to achieve a flat light curve in the region of the primary eclipse. 
The data around the primary and secondary eclipses were analysed using {\sc jktebop} version 43 \citep{2010MNRAS.408.1689S}.\footnote{\url{https://www.astro.keele.ac.uk/jkt/codes/jktebop.html}} 
We included a linear polynomial assigned to scale the model fluxes for each observed eclipse in the model with coefficients included as free parameters in the least-squares fit. 
Other free parameters in the least-squares fit are: the sum of the stellar radii in units of the semi-major axis (fractional radii), $r_1+r_2=(R_1+R_2)/a$; the ratio of the stellar radii, $k=R_2/R_1$; the ratio of the surface brightness at the centre of each stellar disc, $J_0$; the orbital inclination, $i$; the time of mid-primary eclipse, $T_0$; the orbital period, $P$;  $e\sin(\omega)$ and $e\cos(\omega)$, where $e$ is the orbital eccentricity and $\omega$ is the longitude of periastron for the primary star. 
The values of $T_0$ and $P$ obtained provide the following linear ephemeris suitable for calculating the phase of mid-primary eclipse for the Julian date range 2458362\,--\,2460208:
\label{sec:ephem}
\begin{equation}
\label{eq:ephem}
{\rm BJD} ~T_{\rm mid}  =    2459100.59349(1) +  24.5921766(4)\,{\rm E}.
\end{equation}
The values of $k$, $r_1+r_2$, $i$, etc. are consistent with the values derived by \cite{2020MNRAS.498..332M}.

\begin{figure}
\begin{center}
\includegraphics[width=0.5\textwidth]{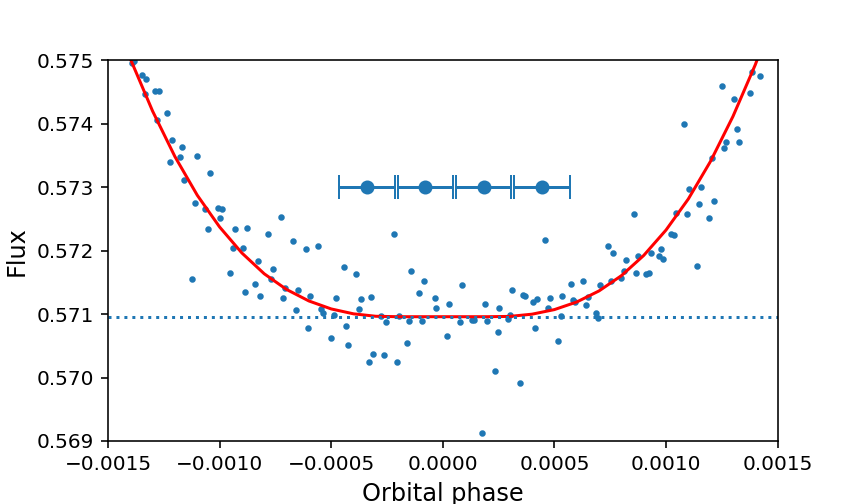}
\end{center}
\caption{TESS photometry of AI~Phe~At orbital phases close to mid-primary eclipse. The flux scale is relative to the mean flux of the binary system out of eclipse. The solid line shows are best-fit model light curve computed with {\sc jktebop}. The 4 spectra observed with UVES where obtained at the orbital phases indicated by points with horizontal error bars.}
\label{fig:PrimaryMinimum}
\end{figure}

The orbital phases computed with this ephemeris covered by the 4 spectra observed with UVES are shown in Fig.~\ref{fig:PrimaryMinimum} compared to the TESS photometry at orbital phases close to mid-primary eclipse and the best-fit model light curve computed with {\sc jktebop}. From the model light curve, we estimate that the contamination in the TESS band of the K0\,IV star spectrum by flux from the F7\,V star is less than 0.014\,per~cent for all 4 spectra and less than 0.002\,per~cent for the two spectra closest to the centre of the eclipse. The flux contamination in the U band will be about 3 times larger than these values, i.e. about 0.07\,~per~cent at the extreme blue end of the average UVES spectrum.

\begin{figure*}
\begin{center}
\includegraphics[width=0.99\textwidth]{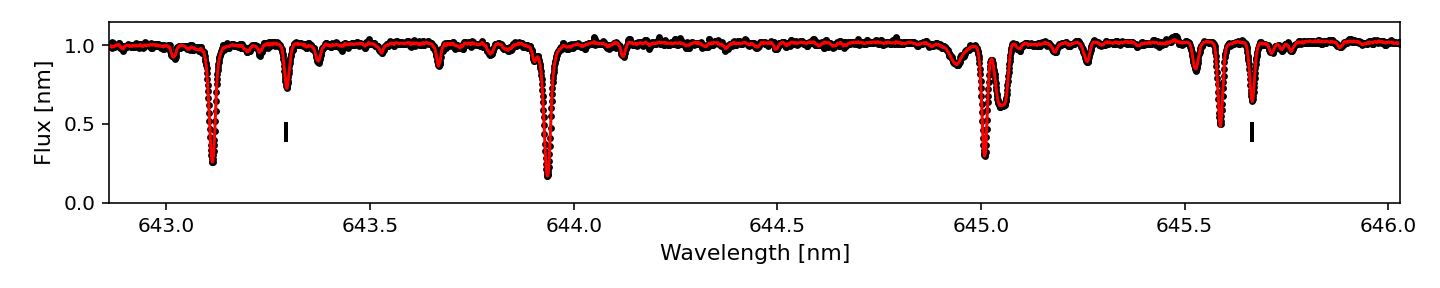}
\includegraphics[width=0.99\textwidth]{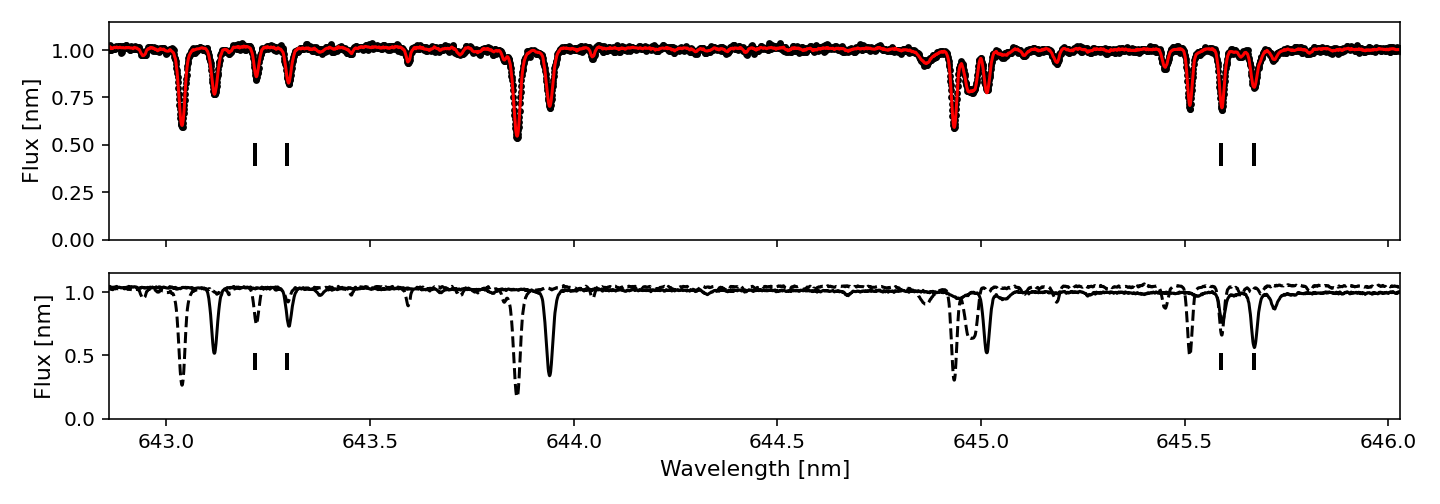}
\end{center}
\caption{Typical spectra of AI~Phe observed with the UVES (upper panel) and HARPS (middle panel) spectrographs. % in the vicinity of the H$\beta$ line. 
The observed data are shown as points and the reconstructed spectra computed from the disentangled component spectra are plotted with red line. 
In the lower panel, the individual component spectra of the K0\,IV star (dashed line) and F7\,V star computed by disentangling the combined spectra are shown at the orbit phase corresponding to the middle panel.
Vertical tick marks indicate Fe\,{\sc ii} lines used in our analysis.}
\label{fig:harps_v_uves}
\end{figure*}

\section{Spectral disentangling}

We used our own implementation of the spectral disentangling algorithm described by \citet{1994A+A...281..286S} to obtain the individual spectra of AI~Phe A and B from the combined HARPS spectra. This algorithm treats the disentangling problem as a matrix equation of the form
\begin{equation}
\label{eq:Mx=b}
\textbfss{M} \bmath{x} = \bmath{b},
\end{equation}
where the vector $\bmath{b}$ contains the observed spectra, the vector $\bmath{x}$ contains the individual component spectra, and the matrix $\textbfss{M}$ sums the two component spectra after a linear interpolation of $\bmath{x}$ onto the wavelength scales of the observed spectra accounting for the Doppler shifts due to the orbital motion of the two stars in the binary system.  
To measure the radial velocities of the two stars we used the cross-correlation algorithm implemented in {\sc iSpec}\footnote{\url{https://www.blancocuaresma.com/s/iSpec}} \citep{2014A&A...569A.111B} with a numerical mask based on the solar spectrum. 
$\textbfss{M}$ is a sparse matrix with a block-diagonal structure so the solution of this equation in the least-squares sense, i.e. the vector  $\bmath{x}$ that minimises $|| \textbfss{M} \bmath{x} - \bmath{b} ||$, can be computed efficiently using the 
LSQR algorithm \citep{1982ACMTM...8...43P}.

The construction of the matrix $\textbfss{M}$ is typically done assuming that the flux ratio between the two component spectra in the observed spectrum is constant. 
In this case, there is an inherent ambiguity in the solution because we can add some arbitrary constant $c$ to the spectrum of one star and subtract the same constant from the spectrum of the other star without changing $\textbfss{M} \bmath{x}$. 
The value of the constant $c$ can be determined by requiring the flux ratio between the spectra to be consistent with the value inferred from the analysis of the light curve. 
However, this only works if the observed spectra can be put onto a consistent flux scale, i.e. normalised such that the flux in the continuum is 1. 
This is not possible for the observed spectra at the far-blue end of wavelength range we observed because there is no continuum visible in the spectrum at these wavelengths. 

To deal with this problem, we modified our implementation of the disentangling algorithm so that we are able to include the UVES spectroscopy obtained during the primary eclipse in the vector of observed spectra, $\bmath{b}$. 
This removes the ambiguity in the flux ratio because the depth of the spectral lines from the K0\,IV star are known from the spectra observed in eclipse with negligible contribution from the F7\,V star. 
This is done by setting the elements of $\textbfss{M}$ corresponding to the contribution of the F7\,V star during eclipse to $0$.
The absorption lines in the two output spectra will then have equivalent widths close to their true values with negligible contamination by the flux from the companion star.   
We also modified the algorithm to weight the spectra in $\bmath{x}$ by their mean signal-to-noise ratio. The corresponding elements of $\textbfss{M}$ are also scaled by the same factor so this amounts to inverse-variance weighting of the observed spectra in the least-squares solution of $|| \textbfss{M} \bmath{x} - \bmath{b} ||$.

Accurate recovery of the individual spectra requires that the flux scale of the spectra is consistent across the entire set of spectra. 
This is difficult to achieve for the full spectral range covered by HARPS so we applied our disentangling software to short segments of the spectra corresponding to the spectral orders produced by the HARPS spectrograph. 
The elements of the vector $\bmath{b}$ are computed by interpolating the HARPS and UVES spectra onto the same wavelength scale that is uniform in logarithmic wavelength with 8192 pixels per order. This sampling of the interpolated spectra was chosen to be approximately 4 times spectral resolution in order to preserve line-profile information as far as possible. 

The disentangling is done in several steps. 
We first apply an approximate continuum normalisation using the algorithm provided in {\sc iSpec} to the spectral segments. 
For spectral segments with wavelengths longer than for 560\,nm we applied a correction for telluric absorption due to H$_2$O and O$_2$ molecules calculated using a fit to each observed spectrum using the software package {\tt TelFit} version 1.4.0 \citep{2014AJ....148...53G}.\footnote{\url{https://telfit.readthedocs.io/}} 
This telluric correction works well enough for the HARPS spectra using the default settings.  
For the UVES spectra, we used the spectrum of the K0\,IV star reconstructed by disentangling without any telluric correction of the input spectra as a model for the source spectrum in the calculation of the telluric correction.
We then disentangle these normalised spectra with uniform weighting to produce a first estimate of the spectra for the two components. 
By reconstructing the observed spectra from these individual spectra we can identify any outliers in the observed spectra and ``patch'' them with the corresponding value from the reconstructed spectrum. 
There are typically only one or two outliers per spectrum in each spectral segment. 
The disentangling is done for a range of assumed flux ratio values so that we can find the optimum flux ratio that gives the lowest root mean square (rms) for the residuals between the observed and reconstructed UVES spectra obtained in eclipse.

We then repeat the unweighted disentangling with these ``patched'' spectra and compute the residuals for each spectrum from its reconstructed spectrum. 
We divide each observed spectrum by a low-order polynomial fit to these residuals. This brings the observed spectra onto a consistent flux scale. 
The standard deviation of the residuals from the polynomial fit is used to compute the mean signal-to-noise ratio of each spectrum.
These signal-to-noise values are used to compute the appropriate weights for each spectrum in the final disentangling step.

HARPS order 92 is only partially covered by the wavelength range of the UVES spectra so for this order we calculated the optimum flux ratio from the overlapping region as for the other orders and then used this value of the flux ratio to disentangle the entire order using only the HARPS spectra. There is no overlap between the UVES spectra and HARPS order 91 so for this order we used only the HARPS spectra for the disentangling and extrapolated a linear fit to the flux ratio versus order number for orders 95 to 117 to infer a flux ratio of $1.256 \pm 0.011$ for this order. 

To combine the short segments of spectrum for each star into a single spectrum, we multiply each section by a linear function such that the mean value in the overlapping sections at the end of each section are the same. This is done in five wavelength regions determined by the gaps in the UVES and HARPS spectra as follows:
379.0\,--\,449.1\,nm ({\it violet});
464.5\,--\,530.4\,nm ({\it blue});
533.8\,--\,558.9\,nm ({\it green});
568.6\,--\,615.9\,nm ({\it yellow});
616.6\,--\,663.0\,nm ({\it red}).
Some of segments show a noticeable curvature in the opposite sense between the two stars. 
This implies a rapid change in flux ratio that is not realistic, so this is likely to be an artifact of the disentangling algorithm.
To remove this curvature, we perform a robust fit of a quadratic function to the difference between the disentangled spectra of star A and star B. 
We then subtract this quadratic function divided by a factor 2 from the  disentangled spectra of star A and add the same function to the spectrum of star B.  
There is some structure in the residuals around strong spectral features in the violet region with a peak-to-peak amplitude of a few per~cent. 
We ascribe this to inconsistencies between the scattered light correction between the HARPS and UVES spectra or, perhaps, inaccurate sky subtraction for some of the HARPS spectra.
These disentangled spectra are available in the supplementary online information that accompanies this article and from the corresponding author's website.\footnote{\url{https://www.astro.keele.ac.uk/pflm/BenchmarkDEBS/}} 
An extract of typical spectra observed with the UVES and HARPS spectrographs are shown in Fig.~\ref{fig:harps_v_uves}.
The signal-to-noise ratio of the disentangled spectra is approximately 250 near 650\,nm.

\section{Analysis}
\label{sec:analysis}

The disentangled spectra were re-normalised using the \texttt{SUPPNET} normalisation tool\footnote{\url{https://github.com/RozanskiT/suppnet}} \citep{Rozanski2022} for each wavelength range. We adopted stellar parameters from Table~\ref{tab:stellar_properties} and used a microturbulence parameter $\vmic = 1.5$\,km~s$^{-1}$ for AI~Phe~A and $\vmic = 1.0$\,km~s$^{-1}$ for AI~Phe~B. These are typical microturbulence values for stars in the respective $T_{\rm eff}$ range \citep{2026A&A...705A.167C} and they are kept fixed mainly to make the two analyses presented below more directly comparable. 
Observing the same set of spectral lines in both stars is advantageous from the point of view of comparing like with like. Line-specific biases (in particular those stemming from atomic data) effectively cancel when determining line-by-line abundance differences. However, as the two stars have different stellar parameters, the same lines will have significantly different line strengths. This adds a certain microturbulence dependence to the analysis. After presenting our analyses, we address this point explicitly.

\subsection{PySME/webSME}
\label{subsec:pysme}

We analysed a set of seven \mbox{Fe\,{\sc ii}} lines, four from the Gaia-ESO line list \citep{Heiter2021} with quality flags Y or Y|U,  complemented by three lines from \citet{Korn2003}. 
These lines cover line strength ranging from 39 to 80 m\AA\ for AI~Phe~A and from 34 to 65 m\AA\  for AI~Phe~B. 
The server-based version of pySME (webSME\footnote{\url{https://websme.chetec-infra.eu/}}, Puschnig et al., in prep.) was used as an analysis interface to the 1D radiative-transfer and fitting code of SME \citep{2017A&A...597A..16P}.
Table \ref{tab:line_abundances} lists the derived line abundances. 
The $\log(gf)$ values used for the analysis of Fe and Mg lines with  pySME and TSFitPy are all taken from \citet{Heiter2021}. The differential analysis conduced with {\sc q2} is independent of the $\log(gf)$ values used.

\begin{table*}
\caption[]{1D LTE abundances and abundance differences measured from individual \mbox{Fe\,{\sc ii}} lines using pySME, TSFitPy and {\sc q2}. log A(Fe) = log ($n$(Fe)/$n$(H)) + 12. The {\sc q2} analysis is line-by-line differential to the Sun and thus reported as a logarithmic abundance relative to the solar abundance: \mbox{[Fe/H]$_{\rm X}$ = log~A(Fe)$_{\rm X} - $log~A(Fe)$_\odot$}.}
\label{tab:line_abundances}
\begin{center}
  \begin{tabular}{@{}lccccccccccc}
%   \begin{tabular}{@{}lrrrrrrrrrrr}
\hline
  \multicolumn{1}{@{}l}{Line} &
 \multicolumn{3}{c}{pySME} &~&
 \multicolumn{3}{c}{TSFitPy} &~& \multicolumn{3}{c}{\sc q2}\\
 \multicolumn{1}{@{}c}{[\AA]} & 
 \multicolumn{1}{c}{log A(Fe)$_{\rm A}$} &
 \multicolumn{1}{c}{log A(Fe)$_{\rm B}$} & 
 \multicolumn{1}{c}{A$-$B} & &
  \multicolumn{1}{c}{log A(Fe)$_{\rm A}$} &
 \multicolumn{1}{c}{log A(Fe)$_{\rm B}$} &
 \multicolumn{1}{c}{A$-$B} & &
  \multicolumn{1}{c}{[Fe/H]$_{\rm A}$} &
 \multicolumn{1}{c}{[Fe/H]$_{\rm B}$} &
 \multicolumn{1}{c}{A$-$B} \\
\hline
 5264.8     &  7.31 & 7.47 & $-0.16 $&& 7.30 & 7.44 &$-0.14 $ && $-0.17 $ & $-0.05 $ & $-0.12 $ \\ 
 5284.1     &  7.28 & 7.42 & $-0.14 $&& 7.30 & 7.41 &$-0.11 $ && $-0.21 $ & $~~0.00 $ & $-0.21 $ \\ 
 5425.2     &  7.23 & 7.35 & $-0.12 $&& 7.21 & 7.30 &$-0.09 $ && $-0.20 $ & $-0.01 $ & $-0.19 $ \\ 
 5991.3     &  7.41 & 7.52 & $-0.11 $&& 7.36 & 7.38 &$-0.02 $ && $-0.12 $ & $-0.11 $ & $-0.01 $ \\ 
 6247.5     &  7.36 & 7.49 & $-0.13 $&& 7.26 & 7.45 &$-0.19 $ && $-0.13 $ & $-0.04 $ & $ -0.09 $ \\ 
 6432.6     &  7.27 & 7.36 & $-0.09 $&& 7.26 & 7.29 &$-0.03 $ && $-0.15 $ & $-0.03 $ & $-0.12 $ \\ 
 6456.3     &  7.34 & 7.41 & $-0.07 $&& 7.34 & 7.37 &$-0.03 $ && $-0.15 $ & $ -0.02 $ & $-0.13 $ \\ 
\hline
Mean &     $7.314$ &     $7.431$ &    $-0.117$ &&     $7.290$ &      $7.377$ &    $-0.087$ &&    $-0.161$ &    $-0.037$ &    $-0.124$ \\
 $\pm \sigma$    & $\pm 0.023$ & $\pm 0.024$ & $\pm 0.012$ && $\pm 0.019$ & $\pm 0.024$  & $\pm 0.024$ && $\pm 0.013$ & $\pm 0.014$ & $\pm 0.025$ \\
\hline
\end{tabular}
\end{center}
\end{table*}

Continuum placement, macroturbulence and line abundance are free parameters in the fitting procedure. 
To account for blending lines, we adopted the Gaia-ESO atomic line list \citep{Heiter2021}.  LTE is assumed, a valid assumption for the stars we study in this paper.
The metallicity for the MARCS model atmospheres \citep{Gustafsson2008} used here with solar abundances from  \citet{Asplund2021} was harmonized with the derived mean abundance. 
Using the full atomic and molecular line list of Gaia-ESO, we verify that the derived abundance difference is not an artefact of line-list incompleteness.
Using the full line list, the abundance difference decreases by no more than 0.01\,dex. 

Given the precise and accurate stellar parameters (in particular with respect to $\log g$ to which the strength of singly ionized lines are sensitive), the derived abundance difference ($-$0.117 $\pm$ 0.012 dex) between the two components is likely physically real. 
\mbox{Fe\,{\sc ii}} lines form fairly deep in the atmosphere (under near-LTE conditions) where 3D effects are relatively small \citep{2019A&A...630A.104A} so an analysis conducted using 3D model atmospheres is unlikely to null this result (see Sect. \ref{subsec:xidep} for a more quantitative assessment). 
Molecular absorption from blending lines could be stronger in 3D and lower the iron abundances of rather weak lines in the B component. 
However, this is  unlikely to change the abundances derived from the stronger lines also used in this work. 
A full analysis using 3D model atmospheres is deferred to a separate paper. 

Magnesium abundances are derived from three weak lines in the red part of the optical spectrum: \mbox{Mg\,{\sc i}} 6318.7\,\AA, \mbox{Mg\,{\sc i}} 6319.2\,\AA\ and \mbox{Mg\,{\sc i}} 6319.4\,\AA. 
This triplet of high-excitation lines is fitted in one go to derive the mean magnesium abundance for each star. 
The PySME least-squares fitting errors are reported as the standard error on these measurements. 
We derive log A(Mg)$_{\rm A} = 7.55 \pm 0.06$ and log~A(Mg)$_{\rm B} = 7.66 \pm 0.05$. 
This implies a similar abundance difference between the two components as for iron. 
Running the webSME line synthesis in non-local thermodynamic equilibrium (NLTE) would lower both abundances by 0.01\,dex leaving the abundance difference unchanged.  

In conclusion, using webSME we derive small but systematic abundance differences which are further elucidated by independent spectroscopic analyses in the following subsections.  

\subsection{TSFitPy}
\label{subsec:tsfitpy}

\begin{figure}
\begin{center}
\includegraphics[width=0.45\textwidth]{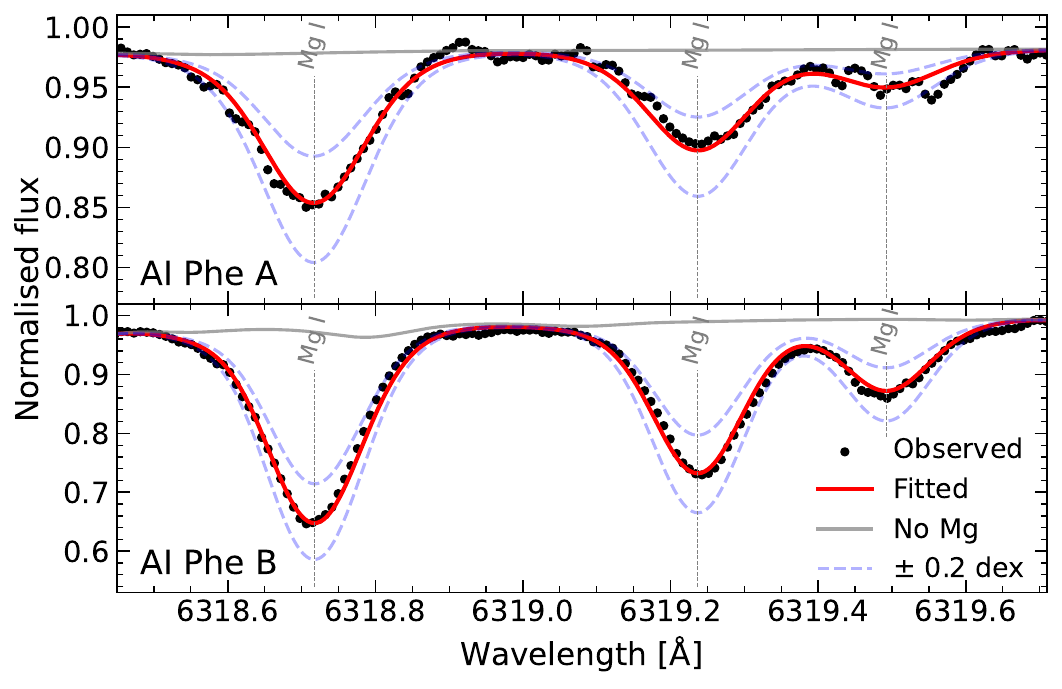}
\end{center}
\caption{Fits of Mg I lines in each of the AI Phe components using the \texttt{TSFitPy} code. This region of the spectrum is affected by some strong telluric absorption features that produce some additional noise in the disentangled spectra.}
\label{fig:mg_tsfitpy_fit}
\end{figure}

Iron and magnesium abundance values were also determined using the {\sc Turbospectrum NLTE} radiative transfer code and its {\sc TSFitPy} wrapper tool \citep{Gerber2023, Storm2023}. 
We adopted the 1D MARCS model atmospheres \citep{Gustafsson2008} and, for the closest internal comparison possible, \citet{Asplund2021} solar abundances. 
The same \mbox{Fe\,{\sc ii}} lines as in webSME were analysed in LTE.
For these lines, the normalisation was not adjusted during fitting and $\vmac$ was freely fitted for each individual line. The choice of keeping the continuum fixed explains the small offsets relative to the results derived with webSME (falling within the mutual 1$\sigma$ error limits).
 
For \mbox{Mg\,{\sc i}}, we used the weaker lines in the region 6318-6319\,\AA\ (the same lines as in the webSME analysis), which are weakly sensitive to NLTE effects. 
However, during the fitting it was noticed that the continuum is visibly offset by around 0.01\,--\,0.02 normalised flux units (abundance difference of $\approx 0.1$ dex). 
Thus an additional option was added in \texttt{TSFitPy} to simultaneously fit the continuum during the fitting procedure by adjusting it using a linear function. 
Therefore, log~A(Mg), $\vmac$ and continuum were freely fitted without any prior restrictions, resulting in 
log~A(Mg)$_{\rm A}$ = 7.47 $\pm$ 0.06  and log~A(Mg)$_{\rm B}$  = 7.63 $\pm$ 0.02.
The corresponding fits are plotted in Fig. \ref{fig:mg_tsfitpy_fit}.

\subsection{Differential analysis with {\sc q2}}
\label{subsec:q2}

\begin{table}
\caption[]{Equivalent widths used in the differential analysis with {\sc q2}, followed by their respective abundances.} 
\label{tab:q2_eqw}
\begin{center}
  \begin{tabular}{@{}lcccccccccc}
\hline 
  \multicolumn{2}{@{}l}{Line [\AA]} &
 % \multicolumn{3}{c}{} &~&
 % \multicolumn{1}{c}{ } &
 \multicolumn{3}{c}{EW [m\AA]} &
  \multicolumn{2}{c}{[Fe/H]} \\
 \multicolumn{2}{@{}c}{} & 
 \multicolumn{1}{c}{A} &
 \multicolumn{1}{c}{B} &
 \multicolumn{1}{c}{Sun}&
 \multicolumn{1}{c}{A} &
 \multicolumn{1}{c}{B }\\
\hline 
\noalign{\smallskip}
{\it Case 1} \\
4779.44   &  Fe {\sc i}     &    23.8    &             &    41.3 & $-0.14$ & - \\
5054.64   &  Fe {\sc i}     &    21.3    &             &    40.6 & $-0.20$ & - \\
5127.68   &  Fe {\sc i}     &    4.8     &             &    21.4 & $-0.24$ & - \\
5295.31   &  Fe {\sc i}     &    17.3    &    40.9     &    30.0 & $-0.12$ & $-0.08$ \\
5386.33   &  Fe {\sc i}     &    17.5    &    46.1     &    33.1 & $-0.16$ & $-0.08$ \\
5466.99   &  Fe {\sc i}     &    20.0    &             &    35.6 & $-0.11$ & - \\
5522.45   &  Fe {\sc i}     &    25.1    &             &    44.1 & $-0.18$ & - \\
5546.51   &  Fe {\sc i}     &    33.7    &             &    52.3 & $-0.14$ & - \\
5560.21   &  Fe {\sc i}     &    33.8    &             &    52.4 & $-0.14$ & - \\
5577.02   &  Fe {\sc i}     &    5.8     &    17.0     &    12.2 & $-0.19$ & $-0.09$ \\
5696.09   &  Fe {\sc i}     &    6.5     &    21.3     &    13.3 & $-0.16$ & $-0.07$ \\
5705.47   &  Fe {\sc i}     &    22.1    &    51.4     &    38.9 & $-0.16$ & $-0.06$ \\
5778.45   &  Fe {\sc i}     &    8.9     &    45.6     &    22.3 & $-0.16$ & $-0.06$ \\
5784.66   &  Fe {\sc i}     &    11.4    &    45.8     &    27.6 & $-0.21$ & $-0.07$ \\
5793.91   &  Fe {\sc i}     &    18.5    &    46.4     &    33.1 & $-0.14$ & $-0.06$ \\
5809.22   &  Fe {\sc i}     &    27.4    &             &    50.4 & $-0.20$ & - \\
5852.22   &  Fe {\sc i}     &    24.0    &             &    40.6 & $-0.15$ & - \\
5855.08   &  Fe {\sc i}     &    11.7    &    32.4     &    22.6 & $-0.16$ & $-0.07$ \\
5905.67   &  Fe {\sc i}     &    39.9    &             &    58.7 & $-0.13$ & - \\
5927.79   &  Fe {\sc i}     &    26.0    &    50.8     &    45.4 & $-0.19$ & $-0.14$ \\
6005.54   &  Fe {\sc i}     &    9.9     &    45.4     &    22.7 & $-0.12$ &  $-0.06$ \\
6271.28   &  Fe {\sc i}     &    11.7    &    43.5     &    24.3 & $-0.12$ & $-0.05$ \\
6380.74   &  Fe {\sc i}     &    34.0    &             &    52.5 & $-0.13$ & - \\
6392.54   &  Fe {\sc i}     &    7.5     &    40.6     &    17.8 & $-0.09$ & $-0.08$ \\
6597.56   &  Fe {\sc i}     &    28.8    &             &    44.1 & $-0.11$ & - \\
6625.02   &  Fe {\sc i}     &    4.7     &    45.8     &    15.4 & $-0.13$ & $-0.09$ \\
5414.07   &  Fe {\sc ii}    &    34.7    &    28.1     &    26.9 & $-0.13$ & $-0.03$ \\
6369.46   &  Fe {\sc ii}    &    23.9    &    19.1     &    18.8 & $-0.14$ & $-0.08$ \\
\noalign{\smallskip}
{\it Case 2}\\
5264.80   &  Fe {\sc ii}    &    55.1    &    43.7     &    45.9 & $-0.17$ & $-0.05$ \\
5284.10   &  Fe {\sc ii}    &    70.8    &    62.3     &    62.9 & $-0.21$ & $\ \ 0.00$ \\
5425.25   &  Fe {\sc ii}    &    49.6    &    42.1     &    42.5 & $-0.20$ & $-0.01$ \\
5991.37   &  Fe {\sc ii}    &    40.9    &    34.3     &    32.8 & $-0.12$ & $-0.11$ \\
6247.56   &  Fe {\sc ii}    &    67.4    &    45.8     &    53.0 & $-0.13$ & $-0.04$ \\
6432.68   &  Fe {\sc ii}    &    51.1    &    40.8     &    41.2 & $-0.15$ & $-0.03$ \\
6456.38   &  Fe {\sc ii}    &    76.9    &    54.9     &    63.0 & $-0.15$ & $-0.02$ \\
\noalign{\smallskip}
\hline
\end{tabular}
\end{center}
\end{table}

In order to validate the findings shown in the Sect. \ref{subsec:pysme} and \ref{subsec:tsfitpy}, we discuss iron abundances for AI~Phe~A and B calculated using a differential line-by-line analysis method, as detailed in \cite{bedell/14}.

We defined the Sun as a reference star, as observed with the HARPS spectrograph. We measured the equivalent widths of \mbox{Fe\,{\sc i}} and \mbox{Fe\,{\sc ii}} lines in the same way in each spectrum  with {\sc iraf} \citep{doug/86}, and then with the aid of the 2019 version of the 1D radiative transfer code {\sc moog} \citep{sneden/73}, {\sc q2} code \citep{ramirez/14} and the grid of standard {\sc marcs} model atmospheres \citep{Gustafsson2008} we calculated the 1D LTE Fe abundances for both stars relative to the Sun. The line list adopted in this run is based on the one from \cite{bedell/18}, but this time excluding lines with strength greater than 80 m\AA. In total, 26 \mbox{Fe\,{\sc i}} lines and 2 \mbox{Fe\,{\sc ii}} lines were considered for the abundance analysis of AI~Phe~A and  14 \mbox{Fe\,{\sc i}} lines  and 2 \mbox{Fe\,{\sc ii}} lines for AI~Phe~B. 

For completeness, the stellar parameters $\teff$, $\logg$, $\vmic$ and [Fe/H] were also calculated through spectroscopic equilibria, where the aim is to achieve excitation and ionization equilibrium balance  between \mbox{Fe\,{\sc i}} and \mbox{Fe\,{\sc ii}} lines \citep{melendez/14}. 
For AI~Phe~A the following stellar parameters were found $\teff=6219\pm63$ K, $\logg=3.98\pm0.11$, $\vmic=1.6\pm0.5~ \rm{km\, s}^{-1}$, and [Fe/H]$=-0.15\pm0.04$.
For AI~Phe~B the method yields the estimates $\teff=5080\pm26$ K, $\logg=3.54\pm0.05$, $\vmic=1.0\pm0.1$ km s$^{-1}$, and [Fe/H]$=-0.09\pm0.03$. 
This is in good agreement with the parameters adopted in this work (see Table \ref{tab:stellar_properties}), despite the reduced number of Fe lines selected here.

On the other hand, if we use the same equivalent widths and fix the stellar parameters from Table \ref{tab:stellar_properties}, with $\vmic = 1.5$\,km~s$^{-1}$ for AI~Phe~A and $\vmic = 1.0$\,km~s$^{-1}$ for AI~Phe~B, we find the following iron abundances: [Fe/H]$_{\rm{A}} = -0.15\pm0.01$ and [Fe/H]$_{\rm{B}} = -0.08\pm0.01$. We notice excellent internal agreement for both stars. The equivalent widths and respective iron abundances for this analysis are identified as {\it Case 1} in Table \ref{tab:q2_eqw}. 1D NLTE corrections based on \cite{amarsi/22} were calculated for Fe {\sc i} lines in both stars using the NLTE abundance-correction tool built into webSME. Absolute corrections vary from $-0.01$ to $0.02$ for AI~Phe A (with an average correction of $0.01$), and from $-0.04$ to $0.00$ for AI~Phe B (with a average correction of $-0.01$). 
Using the 1D NLTE corrections from \cite{bergemann/12}\footnote{\url{https://nlte.mpia.de/gui-siuAC_secE.php}} on the same set of Fe {\sc i} lines, for AI~Phe A the absolute corrections go from $0.00$ to $0.03$ (with an average correction of $0.01$), and from $0.00$ to $0.01$ for AI~Phe B (with an average correction of $<0.01$). The {\em differential} corrections are even smaller, especially for AI~Phe B.
We therefore conclude that -- within the uncertainties of the analysis -- 1D NLTE effects in the differential analysis presented here reduce the abundance difference between the two components by 0.01-0.02 dex.

For the sake of code comparison, we also estimated iron abundances for both stars using {\sc q2} and the same set of \mbox{Fe\,{\sc ii}} lines used with webSME and TSFitPy. 
The equivalent widths and respective iron abundances for this analysis are identified as {\it Case 2} in Table \ref{tab:q2_eqw}. The derived abundances are [Fe/H]$_{\rm{A}} = -0.16\pm0.03$ and [Fe/H]$_{\rm{B}} = -0.04\pm0.03$, which are in good agreement with the determinations using the {\it Case 1} line selection discussed earlier in this subsection. The differential line-by-line abundances are also  presented in Table \ref{tab:line_abundances}.

In conclusion, we find in this analysis AI~Phe~B to be more metal-rich than AI~Phe~A regardless of the set of lines adopted.
The iron-abundance errors we report in this subsection are estimated by considering the stellar-parameter uncertainties and the statistical error associated with the abundances inferred from line-to-line scatter.

\subsection{Dependence of results on modelling assumptions}
\label{subsec:xidep}

One concern in the interpretation of the results given in Table~\ref{tab:line_abundances} is the possibility that the abundance difference for Mg and Fe between the binary components might be artifacts of assumptions made in the modelling, in particular regarding the one-dimensional (1D) stratification of the model atmospheres used and the free parameter microturbulence stemming from the absence of convective motions in these 1D models. In this section we address these concerns.

Grids of 3D corrections to 1D abundances have been presented by e.g.\ \cite{2019A&A...630A.104A}. We have applied these corrections to the lines listed in Table \ref{tab:line_abundances}. The corrections are positive changing abundances by up to 0.12\,dex for individual lines. The metallicities of both stars are shifted to higher  values in 3D. However, the  average correction in the abundance difference between A and B is found to be much smaller (0.02\,dex) lowering the abundance difference in the webSME analysis to ${\rm [Fe/H]}_{\rm A} - {\rm [Fe/H]}_{\rm B} = -0.097\pm 0.012$. As the correction explicitly considers the microturbulence used in the derivation of the 1D abundances, this result should be reliable and trustworthy from a 3D perspective. 

The surface-gravities of the two stars are exceptionally well-constrained (see Table \ref{tab:stellar_properties}). This opens up the possibility of using the wings of strong spectral lines to derive the elemental abundance of the line under study. The wings are pressure-broadened, i.e., collisionally dominated which means they form in LTE. Microturbulence only affects the line core. We evaluated the wings of Mg Ib 5183 using webSME and conclude that they point to an abundance difference of $\approx$0.1\,dex, in good agreement with the abundances derived from the weaker Mg lines in the red part of the spectrum (albeit at 0.13 dex lower absolute abundances).  

A more comprehensive analysis of more elements and detailed comparisons with predictions from stellar-evolution calculations are deferred to a future paper.

%__________________________________________________ MASS RADIUS
\begin{table}
\caption[]{Properties of AI~Phe~A and AI~Phe~B. 
Surface gravity, $g$, is given in cgs units.
References are as follows: 1 -- \citet{1988A&A...196..128A}; 2 -- \citet{2020MNRAS.498..332M}; 3 -- \citet{2020MNRAS.497.2899M}.  
The standard errors given for T${\rm eff}$ are the sum of the random and systematic errors quoted by \citet{2020MNRAS.497.2899M}.
Note that the errors on [Fe/H] are correlated and so the difference in [Fe/H] between AI~Phe~A and AI~Phe~B is more significant that it appears from the difference between the values given in this Table. 
\label{tab:stellar_properties}}
\label{MassRadiusTable}
\begin{center}
  \begin{tabular}{@{}lccc}
\hline
 \multicolumn{1}{@{}l}{Parameter} &
 \multicolumn{1}{c}{AI~Phe~A} &
 \multicolumn{1}{c}{AI~Phe~B} & Ref.\\
\hline
Spectral type & F7 V & K0 IV & 1 \\
Mass [$M_{\odot}$] & $1.194 \pm 0.001$ & $1.244 \pm 0.001$ &  2 \\   
Radius [$R_{\odot}$] & $1.805 \pm 0.002$ & $2.933 \pm 0.002$ & 2 \\   
$\log g$   & $4.002 \pm 0.001$ & $3.598 \pm 0.001 $ &  2 \\
$T_{\rm eff}$ [K] & $6199 \pm 46 $ & $ 5094 \pm 36$ & 3 \\
{[Fe/H]} & $-0.16 \pm 0.05$ & $ -0.08 \pm 0.05 $ & This work. \\
{[Mg/H]} & $-0.04 \pm 0.05$ & $ +0.10 \pm 0.05 $ & This work. \\
\hline
\end{tabular}
\end{center}
\end{table}

\subsection{Adopted values}
 To calculate [Fe/H] and [Mg/H] from the abundances in Table~\ref{tab:line_abundances} we use the reference solar abundance values from \citet{Asplund2021} (log~A(Fe)$_\odot = 7.46 \pm 0.04$, log~A(Mg)$_\odot = 7.55 \pm 0.03$). 
These reference values for the solar abundances have been selected from the various estimates that are available for consistency with the assumed solar abundances used in our analysis, and consistency (within 0.01 dex) with the values of  log~A(Fe)$_\odot$ and log~A(Mg)$_\odot$ used by  \citet{Souto2019} for their analysis of stars in M67 discussed below. 
The mean value of our two or three estimates for [Fe/H] and [Mg/H] are given in Table~\ref{tab:stellar_properties}.
There is no simple way to accurately estimate the standard error of these mean values since they are based on only two or three values that are not independent from one another.
It is not likely that our estimates of these abundances are more precise than the solar reference values, so we can take the standard errors on the solar reference values as a lower limit to the standard errors on our abundance estimates.
Guided by this and the standard errors on the individual measurements, we assign an estimated standard error of 0.05 to all our abundance estimates. 

\section{Discussion}
\begin{figure}
\begin{center}
\includegraphics[width=0.45\textwidth]{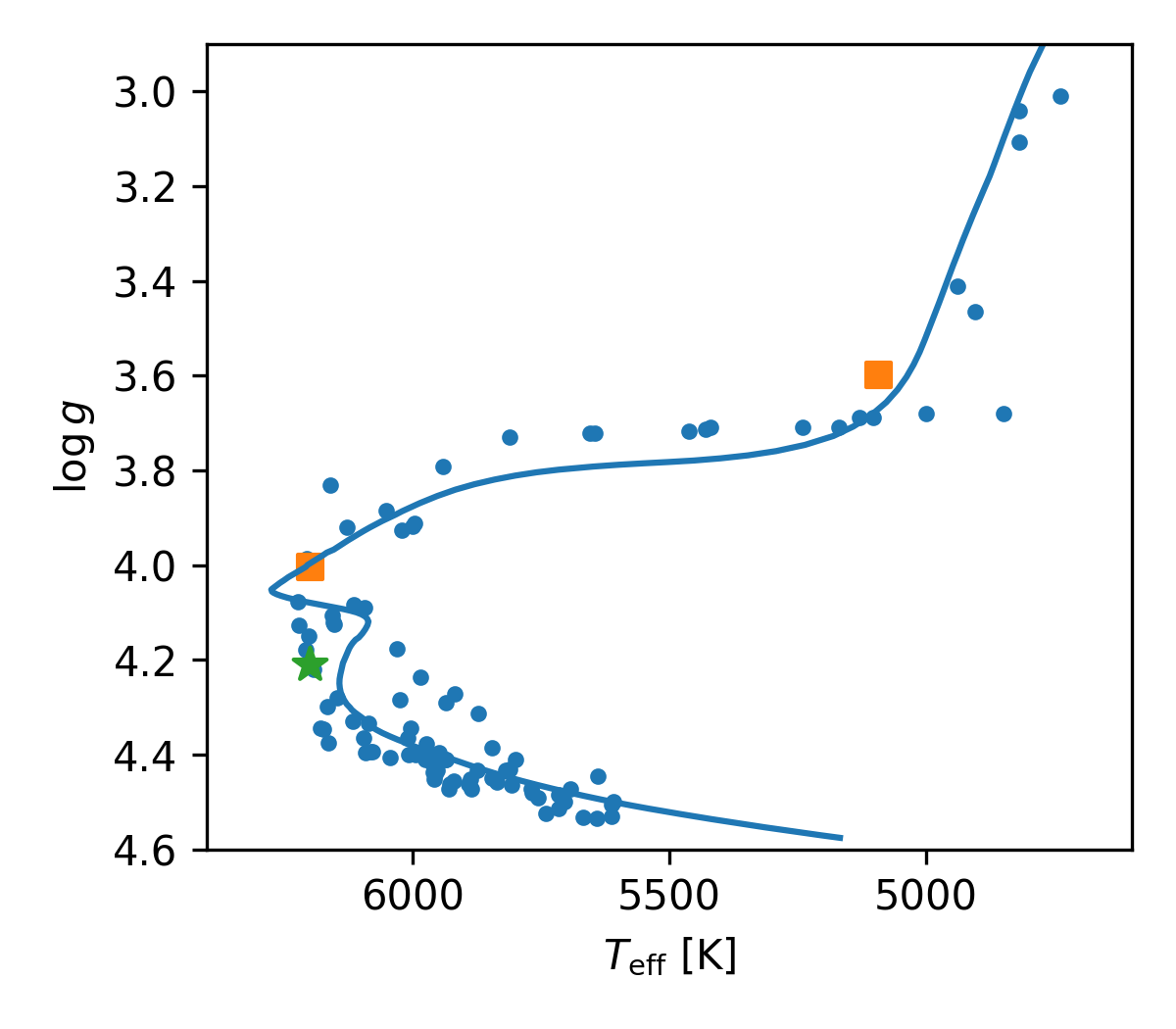}
\end{center}
\caption{AI Phe in the Kiel diagram (squares) compared to stars in the open cluster M67 (dots). ``Isochrone A'' from \protect{\citet{2024MNRAS.532.2860R}} with an age of 3.95\,Gyr is shown for context. The star symbol mark the position of WOCS~11028~A.}
\label{fig:M67_Kiel_AI_Phe}
\end{figure}

All analyses presented above imply an iron abundance difference $\approx 0.1$\,dex. 
In particular, the line-by-line abundance difference from the pySME analysis ($-0.097 \pm 0.012$\,dex when corrected for 3D effects) is highly significant (formal significance $> 8\sigma$). 
The properties of the stars in M67 provide useful context for the interpretation of our results, because the age and metallicity of AI~Phe are similar to the age and metallicity of this well known and extensively studied open cluster.
\citet{2017A&A...600A..41V} estimate the age of AI~Phe to be $\approx 4.7$\,Gyr based on a comparison of the properties measured by \citet{2020MNRAS.498..332M} and
\citet{2020MNRAS.497.2899M} to a large grid of stellar models.
Age estimates for M67 are typically $\approx 4$\,Gyr \citep{2010A&A...513A..50B, 2022A&A...665A.126N, 2025MNRAS.538.1720R}, i.e.\ $\lesssim$ 1\,Gyr younger than AI~Phe.
The positions of AI~Phe~A and AI~Phe~B in the Kiel diagram compared to stars in M67 are shown in Fig.~\ref{fig:M67_Kiel_AI_Phe}.
The values of $T_{\rm eff}$ and $\log g$ for stars in M67 in Fig.~\ref{fig:M67_Kiel_AI_Phe} are taken from \citet{2012A&A...541A.150P}.

\begin{figure*}
\begin{center}
\includegraphics[width=0.75\textwidth]{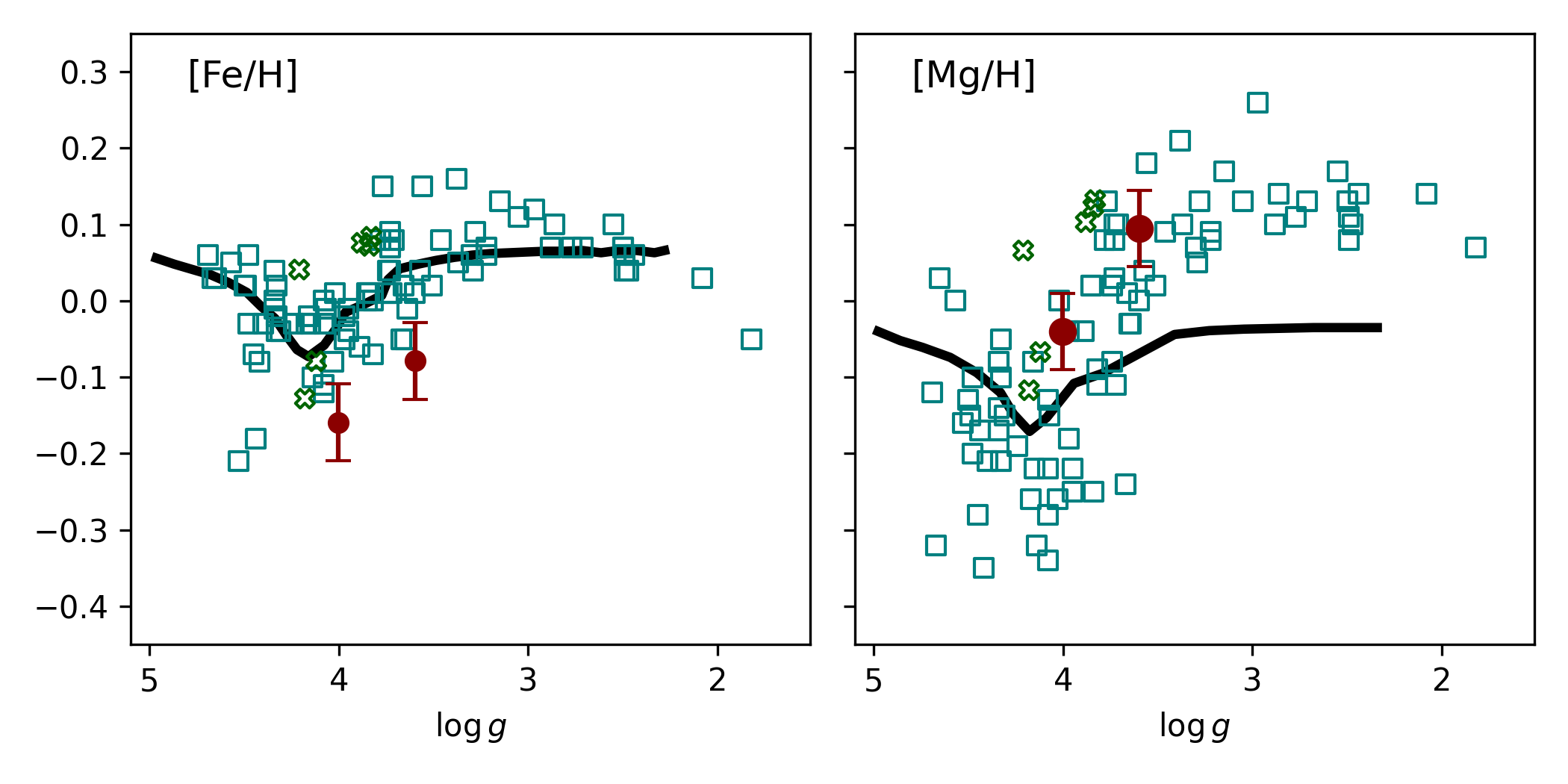}
\end{center}
\caption{The abundance ratios [Fe/H] and [Mg/H] for AI~Phe~A and AI~Phe~B (points with error bars) compared to measurements for stars in M67 from \citet{Souto2019} (open squares) and  \protect{\citet{2019A&A...627A.117L}} (open crosses). Black lines show the abundance ratios predicted by the stellar evolution models including diffusion computed by \citet{Souto2019}.
Note that the errors on [Fe/H] for AI~Phe~A and AI~Phe~B are correlated so the difference in [Fe/H] between the two stars is more significant than it appears to be from the error bars shown here.}
\label{fig:M67_v_AI_Phe_FeH}
\end{figure*}

Our adopted values for [Fe/H] and [Mg/H] for AI~Phe~A and AI~Phe~B are compared to  [Fe/H] and [Mg/H] estimates for stars in M67 as a function of $\log g$ in Fig.~\ref{fig:M67_v_AI_Phe_FeH}. The general pattern is for dwarf stars ($\log g \approx 4$) to have photospheres depleted in metals due to diffusion (gravitational settling) during their lifetimes. 
The abundance of metals then rises again as the stars evolve to become sub-giant stars ($\log g \approx 3.5$) as the deepening convective envelope mixes the metals that have settled into the stellar interior back into the photosphere.
AI~Phe clearly follow this pattern, i.e. the diffusion of elements in the atmospheres of AI~Phe~A and AI~Phe~B is similar to that seen in single stars of M67.
This suggests that AI~Phe is a very suitable benchmark star to calibrate evolution models for single stars including the effects of diffusion.  
This is very clear for [Mg/H] which is impacted more by diffusion than many other elements \citep{Korn2007}, but is also consistent with the relative abundance ratio [Fe/H] for AI~Phe~A and AI~Phe~B.
The magnesium abundance [Mg/H] for AI~Phe is very similar to M67, the iron abundance [Fe/H] is lower by about 0.15~dex than for stars in M67. 
The steep gradient in the abundance ratio [Mg/H] as a function of $\log g$ between dwarf and subgiants that is not matched by the stellar evolution models computed by \citet{Souto2019} for stars in M67 is also seen in AI~Phe.

M67 also contains an eclipsing binary star, WOCS~11028 (Sanders 617), with an orbital period of 62.6 days. 
The properties of this star have been measured by \citet{2021AJ....161...59S}, 
who describe WOCS~11028~A as being at the ``faint turn-off point for the cluster -- a distinct evolutionary point that occurs after the convective core has been established and while the star is in the middle of its movement toward lower surface temperatures, before the so-called hook at the end of the main sequence.''
The position of the primary component, WOCS~11028~A, is also shown in the Kiel diagram in Fig.~\ref{fig:M67_Kiel_AI_Phe}.
The mass of WOCS~11028~A is $1.222\pm0.006 M_{\odot}$, close to the average mass of AI~Phe~A and AI~Phe~B.
\citet{2025MNRAS.538.1720R} have used asteroseismology of the giants and sub-giant stars in M67 to infer an age of $3.95 \pm 0.35$\,Gyr. 
The best-fit isochrones for this age  underestimates the mass of WOCS~11028~A by approximately $0.05 M_{\odot}$. 
A similar issue was seen by \citet{2022A&A...665A.126N} in their analysis of the colour-magnitude diagram of M67.
\citet{2021AJ....161...59S} propose that  ``diffusion is probably necessary to reconcile spectroscopic abundances of M67 stars with the need for higher metallicity models and that reduced strength convective overshooting is occurring for stars at the turn-off.'' 
It would be interesting to test whether these stellar evolution models with 
reduced strength convective overshooting can successfully match the properties of AI~Phe~A and AI~Phe~B assuming the same age and initial composition of the two stars.

\citet{2023A&A...678A.203V} have compared the precise mass, radius and effective temperature measurements for AI~Phe~A and AI~Phe~B to a large grid of stellar evolution models over a range of values for the convective overshooting parameter, $\beta$, and initial helium abundance, $Y$. 
Their study used the value for the photospheric iron abundance of ${\rm [Fe/H]} = -0.14 \pm 0.1$ from \citet{1988A&A...196..128A}, which is consistent with our estimates of [Fe/H] for both stars within the quoted uncertainty. 
Comparison was made to models both with diffusion and without diffusion. 
\citeauthor{2023A&A...678A.203V} found that the range of models without diffusion that are consistent with the observed properties of AI~Phe are mostly confined to those with a convective overshooting parameter in the range $\beta = 0.03$\,--\,$0.07$. 
For models that include diffusion, \citeauthor{2023A&A...678A.203V} found that $\beta$ is ``barely constrained, and found to be in the range of 0.04\,--\,0.12''. 
At first sight, this is a disappointing result because it seems to indicate that the evidence for diffusion in AI~Phe implies that this system is not suitable for calibrating the value of $\beta$ to be used in stellar models. 
However, we note that \citeauthor{2023A&A...678A.203V} set an arbitrary value of 50\,K on the $T_{\rm eff}$ estimates for both stars ``in light of the existing difference in the calibration scale among different literature sources''.
We do not believe that this arbitrary assignment of an error estimate based on a comparison of indirect methods to estimate $T_{\rm eff}$ is warranted for a direct measurement of $T_{\rm eff}$ using the Stefan-Boltzmann law where the sources of systematic error are small and well understood.
In any case, it would certainly be worthwhile to repeat the analysis of \citeauthor{2023A&A...678A.203V} and to perform similar calculations with other model grids to see whether any of these models can match the observed differences in [Fe/H] and [Mg/H] for these exceptionally well-characterised stars.

The spectra we have observed include the Li\,I 6707\,\AA\ absorption line so we are able to estimate the photospheric lithium abundance for AI~Phe~A and AI~Phe~B. 
The abundances we derive are qualitatively in agreement with what is seen in M67 in that AI~Phe~B is depleted in lithium compared to AI~Phe~A, but the lithium abundance of AI~Phe~A is slightly higher than in dwarf stars of M67. 
The interpretation of these photospheric lithium abundances is complicated because it is closely related to the rotational evolution of the stars and internal mixing processes that are not well understood, so we defer a detailed discussion of the lithium abundance to a future study.

\section{Conclusion}
The photospheric abundance ratios for [Fe/H] and [Mg/H] that we have measured for the dwarf and subgiant components of the detached eclipsing binary star AI~Phe clearly show the signature of elemental diffusion (gravitational settling).
This can be recognised from the difference in the values of [Fe/H] and [Mg/H] between the two stars ($\approx -0.1$ dex).
These abundance differences are similar to those observed between dwarf and subgiant stars in the open cluster M67, which have a similar composition and age to AI~Phe.
This makes it possible to use AI~Phe to test stellar evolution models that include elemental diffusion. 
There is also plenty of scope to use the high-quality spectra that we have made available here for two stars that are characterised to very high precision and accuracy to investigate the accuracy of synthetic spectra generated from stellar model atmospheres using radiative transfer codes, and to study the impact of diffusion on other elements.

\section*{Acknowledgements}
The authors thank Anish Amarsi for help with his 3D abundance corrections for iron. 

This research was supported by UK Science and Technology Facilities Council (STFC) research grant numbers ST/R000638/1, ST/Y002563/1 and UKRI1193, and UK Space Agency (UKSA) grant number UKRI966.

A.J.K.\ acknowledges support by the Swedish National Space Agency (SNSA). 

M.B. and M.G. are supported through the Lise Meitner grant from the Max Planck Society. We acknowledge support by the Collaborative Research Centre SFB 881 (projects A5, A10), Heidelberg University, of the Deutsche Forschungsgemeinschaft (DFG, German Research  Foundation) and by the European Research Council (ERC) under the European Union’s Horizon 2020 research and innovation programme (grant agreement 949173).

This research was supported by the Munich Institute for Astro-, Particle and BioPhysics (MIAPbP) which is funded by the Deutsche Forschungsgemeinschaft (DFG, German Research Foundation) under Germany's Excellence Strategy – EXC-2094 – 390783311.

P.J. acknowledges support of ANID FONDECYT Regular Grant Number 1231057. 

%%%%%%%%%%%%%%%%%%%%%%%%%%%%%%%%%%%%%%%%%%%%%%%%%%
\section*{Data Availability}
 The spectra used in this study are available from the European Southern Observatory (ESO) Science Archive Facility (\url{https://archive.eso.org}, programme identifier  0106.D-0165(A)) and are also available from \url{https://www.astro.keele.ac.uk/pflm/BenchmarkDEBS/}. The disentangled HARPS spectra are provided in the supplementary online information and are also available from \url{https://www.astro.keele.ac.uk/pflm/BenchmarkDEBS/}. 

%%%%%%%%%%%%%%%%%%%% REFERENCES %%%%%%%%%%%%%%%%%%

% The best way to enter references is to use BibTeX:

\bibliographystyle{mnras}
\bibliography{all} % if your bibtex file is called example.bib

% Alternatively you could enter them by hand, like this:
% This method is tedious and prone to error if you have lots of references
%\begin{thebibliography}{99}
%\bibitem[\protect\citeauthoryear{Author}{2012}]{Author2012}
%Author A.~N., 2013, Journal of Improbable Astronomy, 1, 1
%\bibitem[\protect\citeauthoryear{Others}{2013}]{Others2013}
%Others S., 2012, Journal of Interesting Stuff, 17, 198
%\end{thebibliography}

%%%%%%%%%%%%%%%%%%%%%%%%%%%%%%%%%%%%%%%%%%%%%%%%%%

%%%%%%%%%%%%%%%%% APPENDICES %%%%%%%%%%%%%%%%%%%%%

%\appendix

%\section{Some extra material}

%%%%%%%%%%%%%%%%%%%%%%%%%%%%%%%%%%%%%%%%%%%%%%%%%%

% Don't change these lines
\bsp	% typesetting comment
\label{lastpage}
\end{document}